\documentclass[aps,prb,twocolumn,superscriptaddress,showpacs,notitlepage,citeautoscript]{revtex4-1}
\usepackage{amsmath}
\usepackage{appendix}
\usepackage{bm}
\usepackage{graphicx}
\usepackage{txfonts}
\usepackage{color}

\newcommand{\bea}{\begin{eqnarray}}
\newcommand{\eea}{\end{eqnarray}}
\newcommand{\bei}{\begin{itemize}}
\newcommand{\eei}{\end{itemize}}
\newcommand{\be}{\begin{equation}}
\newcommand{\ee}{\end{equation}}

\newcommand{\Ef}{E_{\mathrm{F}}}

\newcommand{\ci}{\mathrm{i}}

\begin{document}
\title{Impurities and electronic localization in graphene bilayers}
\author{H. P. Ojeda Collado}
\affiliation{Centro At{\'{o}}mico Bariloche and Instituto Balseiro,
Comisi\'on Nacional de Energ\'{\i}a At\'omica, 8400 S. C. de Bariloche, Argentina.}
\author{Gonzalo Usaj}
\affiliation{Centro At{\'{o}}mico Bariloche and Instituto Balseiro,
Comisi\'on Nacional de Energ\'{\i}a At\'omica, 8400 S. C. de Bariloche, Argentina.}
\affiliation{Consejo Nacional de Investigaciones Cient\'{\i}ficas y T\'ecnicas (CONICET), Argentina.}
\author{C. A. Balseiro}
\affiliation{Centro At{\'{o}}mico Bariloche and Instituto Balseiro,
Comisi\'on Nacional de Energ\'{\i}a At\'omica, 8400 S. C. de Bariloche, Argentina.}
\affiliation{Consejo Nacional de Investigaciones Cient\'{\i}ficas y T\'ecnicas (CONICET), Argentina.}

\begin{abstract}
We analyze the electronic properties of bilayer graphene with Bernal stacking and a low concentration of adatoms. Assuming that the host bilayer lies on top of a substrate, we consider the case where impurities are  adsorbed only on the upper layer. We describe non-magnetic impurities  as a single orbital hybridized with carbon's p$_z$ states. The effect of impurity doping on the local density of states with and without a gated electric field perpendicular to the layers is analyzed. We look for Anderson localization in the different regimes and estimate the localization length. In the biased system, the field induced gap  is partially filled by strongly localized impurity states.  Interestingly, the structure, distribution and localization length of these states depend on the field polarization.
\end{abstract}
\pacs{03.75.Lm,72.25.Dc,71.70.Ej}

\maketitle

\section{Introduction}
Graphene, in all its allotropic forms, is a material with exceptional mechanical, electronic and thermal properties. Its discovery led to one of the most active fields in material science and condensed matter research during the last decade. Graphene monolayer, usually referred simply as graphene, and multilayers have different properties due to a subtle difference in their band structure. 
It is now well established that in graphene monolayers, the electronic excitations with crystal momentum close to the $K$ or $K^{\prime}$ points of the Brillouin zone (BZ), are chiral quasiparticles behaving as massless Dirac fermions. These excitations dominate the low temperature physics  leading to a number of remarkable phenomena in clean samples.\cite{CastroNeto-review,DasSarma2011,Beenakker2008} Impurities, adatoms and structural defects change these properties and there has been a considerable effort to study and characterize the different types of defects and disorder in graphene \cite{Evers2008,Chan2008,Wehling2009,Wehling2010,Wehling2010b,Sofo2012,Roche2012,Matis2012,Guillemette2013,Hong2011} as well as the effect of doping on them\cite{Sofo2011,Chan2011,Guzman2014}.  

The problem of disorder and electron localization has attracted the attention of many groups for Dirac fermions tend to elude localization in systems with Anderson-type disorder.\cite{Aleiner2006,Ostrovsky2006,Ostrovsky2007,Mirlin2010,Konig2012,Gattenloehner2013,Cresti2013,Usaj2014} Impurities leading to short range disorder at the atomic scale generate inter-valley mixing and break the symplectic symmetry opening the route to strong localization.  

Bilayer graphene (BLG) presents some fundamental differences due to its crystallographic structure. It consist of a stacking of two graphene layers and in the most common structure, known as the Bernal stacking, only one of the two non-equivalent sites $(A,B)$ of the honeycomb lattice of the top layer lies on top of a site of the bottom layer. The resulting structure, shown in Fig. \ref{fig1}, induces a weak coupling of the two layers. The unit cell has four carbon atoms leading to four $\pi$-bands, two of them having a parabolic dispersion relation around the $K$ and $K^{\prime}$ points of the BZ and that touch each other at the Fermi energy 
\cite{Nilsson2008,Castro2010,McCann2013}.   

In most of the experimental setups, BLG lies on top of a substrate and the impurities are adsorbed on the top layer only. When atoms like hydrogen or fluorine are adsorbed they are bounded to a single C atom. 

One of the most interesting aspects of this system is that its electronic structure can be controlled with an electric field applied perpendicular to the layers 
\cite{Castro2007, McCann2006, Min2007,Taychatanapat2010}.  In biased BLG a gap opens at the Fermi level and impurities may induce a bound state in the gap \cite{Dahal2008,Mkhitaryan2013}. As noted in Ref. [\onlinecite{Mkhitaryan2013}], the impurity spectral density and the existence of the bound states may depend on the polarity of the field. A finite impurity concentration generates a gate dependent impurity band creating new and encouraging alternatives to control the transport properties. However, in contrast to the important activity in the study of disordered graphene, the problem of BLG with a diluted concentration of adatoms inducing short range potentials has not been investigated in detail. 

In this work we study the problem of a low concentration of impurities  in biased and unbiased BLG. 
We present a model that aims to describe fluorinated BLG, an extension of the model of Ref.  [\onlinecite{Usaj2014}] used to discuss the experiments of Ref. [\onlinecite{Hong2011}]. 

In section II we present the model and revisit the single impurity problem. In section III we describe the numerical methods and present results for the local density of states (LDOS) at the different sites. Section IV includes a discussion of localization and a summary and conclusions are presented in section V.

\section{The Model}
\begin{figure}[tb]
\includegraphics[width=0.95\columnwidth]{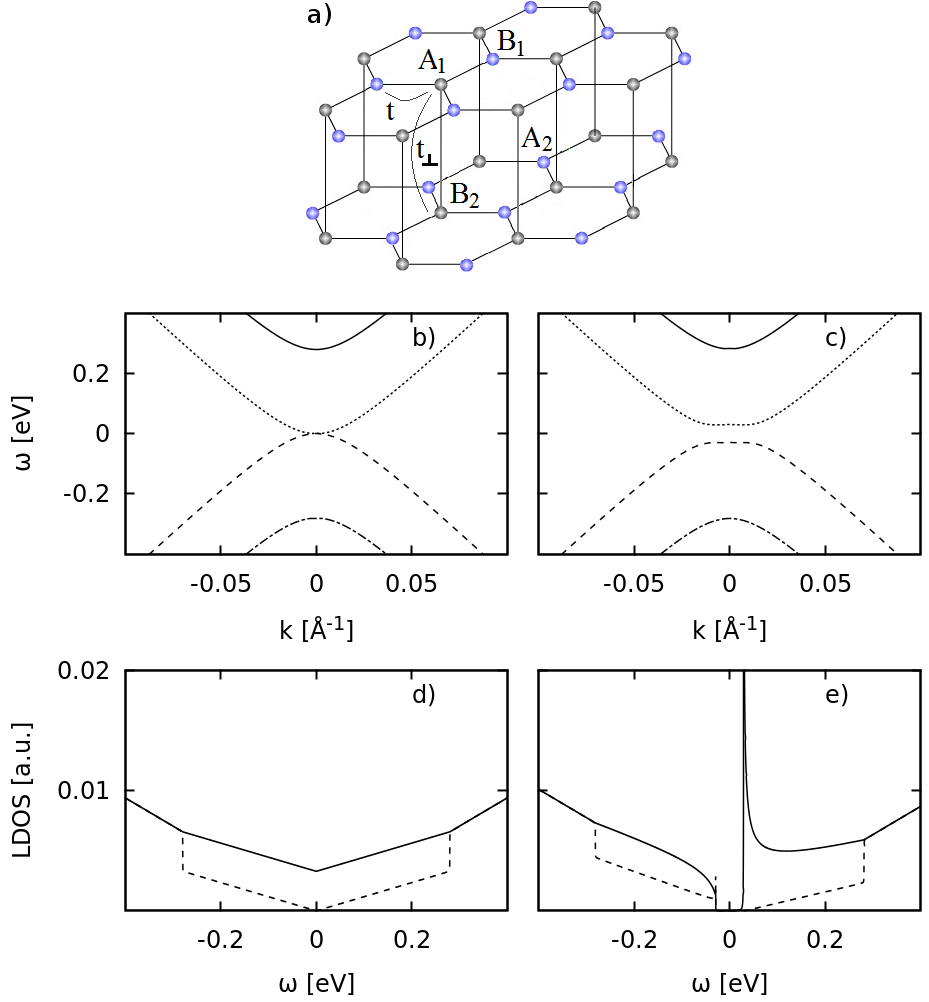}
\caption{(a) Bernal structure of  bilayer graphene. Panels (b) and (c) show the low energy band structure for the unbiased and biased BLG, respectively. In (d) the LDOS for the unbiased case, continuous line corresponds to the $B_{1}$ and $A_{2}$ sites, dashed line to the $A_{1}$ and $B_{2}$ sites. (e) LDOS for a positive ($V>0$) bias voltage. Continuous (dashed) line corresponds to the $B_{1}$ ($A_{1}$) sublattice. For a negative voltage the LDOS can be obtained from the ones with positive $V$ by replacing $\omega$ by $-\omega$.}
\label{fig1}
\end{figure}

The Hamiltonian of the system is $H=H_{\mathrm{BLG}}+H_{\mathrm{imp}}+H_{\mathrm{hyb}}$ where the first term describes the electronic structure of the BLG, the second one is the impurities' Hamiltonian and the last one includes the hybridization between each impurity orbital and the p$_{z}$ orbital of the underlying C atom. In the tight-binding approximation  the BLG Hamiltonian reads
\begin{eqnarray}
\nonumber
H_{\mathrm{BLG}}&=&-\sum_{i,\bm{k},\sigma}V(-1)^{i}\left(a^{\dagger}_{i\bm{k}\sigma}a_{i\bm{k}\sigma}^{}+b^{\dagger}_{i\bm{k}\sigma}b_{i\bm{k}\sigma}^{}\right)\\
\nonumber
&&+\sum_{i,\bm{k},\sigma}t\left(\phi({\bm{k}})\,a^{\dagger}_{i\bm{k}\sigma}b_{i\bm{k}\sigma}^{}+\phi^{*}({\bm{k}})\,b^{\dagger}_{i\bm{k}\sigma}a_{i\bm{k}\sigma}^{}\right)\\
&&-\sum_{\bm{k},\sigma}t_{\perp}\left(a^{\dagger}_{1\bm{k}\sigma}b_{2\bm{k}\sigma}^{}+b^{\dagger}_{2\bm{k}\sigma}a_{1\bm{k}\sigma}^{}\right)\,.
\end{eqnarray}
Here, $a_{i\bm{k}\sigma}$ and $b_{i\bm{k}\sigma}$ destroy electrons with wavevector $\bm{k}$ and spin $\sigma$ in sub-lattices $A$ and $B$, respectively, and the subindex $i=1$($2$)  refers to the top (bottom) plane. $V$ is the bias voltage, $t$ and $t_{\perp}$ are the intra-plane and inter-plane hoppings, respectively, and $\phi(\bm{k})=\sum_{\bm{\delta}}e^{i\bm{k}\cdot\bm{ \delta}}$ where $\{\bm{\delta}\}$ are the three vectors  connecting one site with its neighbors in the same plane. 
In our notation, the $C$ atoms of the top layer in the $A$ sublattice, referred as the $A_{1}$ sublattice,  lie on top of the $C$ atoms in the $B$ sublattice of the bottom layer ($B_{2}$ sublattice), see Fig. \ref{fig1}.  

We describe non-magnetic impurities  as single orbital impurities where the electron-electron interactions are not explicitly included, 
\begin{equation}
H_{\mathrm{imp}}=\sum_{l,\sigma}\varepsilon_0 f^{\dagger}_{l\sigma}f_{l\sigma}^{}\,,
\end{equation}
where $f^{\dagger}_{l\sigma}$ creates an electron on the impurity orbital at site $l$ and energy  $\varepsilon_0$ and the sum runs over the sites of
carbon lattice having an adsorbed impurity on top.  The last term of the
Hamiltonian describes the hybridization of the impurity and
the graphene orbitals of the top layer
\begin{equation}
H_{\mathrm{hyb}}=\gamma\sum_{l\in A_{1},\sigma} (f^{\dagger}_{l\sigma}a_{1l\sigma}^{}+a^{\dagger}_{1l\sigma}f_{l\sigma}^{})+ \gamma\sum_{l\in B_{1},\sigma} (f^{\dagger}_{l\sigma}b_{1l\sigma}^{}+b^{\dagger}_{1l\sigma}f_{l\sigma}^{})\,,\\
\end{equation}
with the sum taken over all the sites with an impurity on top, $a_{1l\sigma}^{}=N^{-\frac{1}{2}}\sum_{\bm{k}}e^{i{\bm{k}}\cdot{\bm{R}}_{l}}a_{1\bm{k}\sigma}^{}$ where ${\bm{R}}_{l}$ is the coordinate of site $l$---a similar expression holds for $b_{1l\sigma}^{}$. Typical values of the microscopic parameters are $t=2.8$ eV and $t_{\perp} = 0.1t$, while the bias voltage $V$ is taken in the range $|V|\le 0.3$ eV and with any loss of generality we take $\varepsilon_0 \ge 0$. With our one electron Hamiltonian, the case  $\varepsilon_0 \le 0$ can be obtained from the previous one by an electron-hole transformation, $\it{i.e.}$ by replacing $\omega$ by $-\omega$ and $V$ by $-V$. In what follows we take $\varepsilon_{0}=0.3$ eV and $\gamma=5.6$ eV.  
Figure \ref{fig1} illustrates the BLG lattice and its band structure.
As we will not consider spin dependent effects, we drop the spin index in what follows.

\begin{figure}[tb]
\includegraphics[width=0.95\columnwidth]{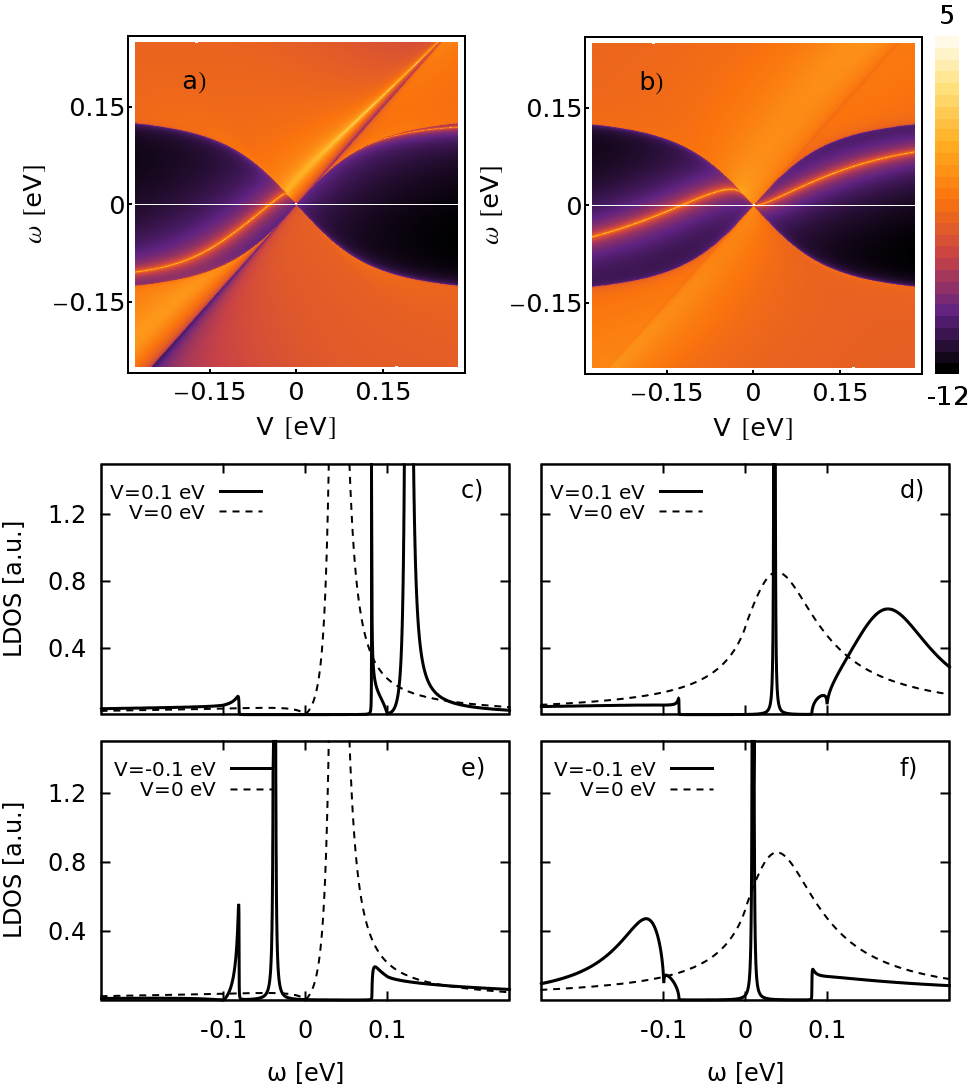}
\caption{Color maps (in logarithmic scale) of the impurity spectral densities in the $[\omega, V]$ plane. (a) and (b) correspond to one impurity on the $A_{1}$ and $B_{1}$ sublattice, respectively. Dark areas indicate the electric field induced gap in the pristine sample. Lower panels show the impurity spectral density for the unbiased (dashed lines) and biased (continuous lines) cases: left and right columns correspond to impurities on the $A_{1}$ and $B_{1}$ sublattices, respectively, and different polarities (as indicated in the insets)}
\label{fig2}
\end{figure}

It is instructive to review some aspects of the single impurity problem before presenting the many impurities case\cite{Dahal2008,Mkhitaryan2013}.
For one impurity, the retarded impurity propagator ${\mathcal{G}}_{ll}= \langle\langle f_{l}^{},f_{l}^{\dag }\rangle\rangle$ takes the form
\begin{equation}
{\mathcal{G}}_{ll}=\frac{1}{\omega+\ci0^+-\varepsilon _{0}-\Sigma (\omega,V)}\,,
\end{equation}
where  $\Sigma (\omega,V)= \gamma^{2}\tilde{g}(\omega,V)$ is the impurity's self-energy and $\tilde{g}(\omega,V)$ is the local propagator of electrons in the C orbital hybridized with the impurity. The carbon-carbon propagator can be evaluated in the continuous limit. Color maps of the impurity spectral densities for impurities on $A_{1}$ and on $B_{1}$ sites are shown in Fig. \ref{fig2}.  

Let us consider first the case of an impurity adsorbed on top of an $A_{1}$ site (left panels of Fig. \ref{fig2}). For $V=0$ the impurity spectral density $\rho_{\mathrm{imp}}(\omega)=-1/\pi\, \mathrm{Im}\mathcal{G}_{ll}$ has the characteristic structure of a resonant state, the real part of the self-energy shifts the maximum from $\varepsilon_{0}\ge 0$ towards the Dirac point generating a narrow resonance at the renormalized energy $\bar{\varepsilon}_0=\varepsilon_0+\mathrm{Re}\Sigma(\bar{\varepsilon}_0,0)$. For the impurity parameters used in this calculation, the renormalized energy is an order of magnitude smaller than the bare energy ($\bar{\varepsilon}_0\ll\varepsilon_0$). Close to the Dirac point the impurity spectral density shows the characteristic $|\omega|$ behavior of the LDOS of the $A_{1}$ sites. For $V\ne 0$ a gap opens at the Dirac point and the impurity spectral density $\rho_{\mathrm{imp}}(\omega)$ may show a bound state within the gap. An important effect of the polarity of the field $V$ is apparent from the figure: a {\it negative} voltage $V$ leads to a bound state within the energy gap close to the top of the valence band while for a {\it positive} $V$ the bound state energy---if observed---lies exponentially close to the conduction band edge. This is due to the structure of the LDOS at the $A_{1}$ sublattice: for positive $V$ the LDOS at the edge of the conduction band $E_{c}$ behaves  as  $\omega-E_{c}$, as in the  4D electron gas where a strong coupling to the impurity is required to split a bound state out of the band (see Fig. \ref{fig1}e)). 

For impurities on the $B_{1}$ sites (right panels of Fig. 2) the results are somewhat different. For $V=0$ the width of the impurity resonance is much broader due to the larger LDOS of the underlying C atom. For small and {\it positive}  $V$ the bound state lies close to the gap centre while for small {\it negative}  $V$ no bound state occurs. This effect can be understand by looking at the LDOS at the $B_{1}$ sites in the biased BLG  (see Fig 1e)). There, the LDOS of the conduction band for small and positive $V$ presents a 1D-like van Hove singularity leading always to a bound state, while for $V<0$ a 3D-like singularity at the edge of the conduction band requires a minimum value of the parameters to split a state out of the band. However, this effect, discussed in Ref. [\onlinecite{Mkhitaryan2013}], is observed only for extremely small values of the bias voltage. For physically relevant values of the gap, bound states occur for both polarities although their position depends on the sign of $V$.

These asymmetries illustrate the importance of the polarity of the electric field on the electronic structure of the impurity doped system. The variation of the impurity energy $\varepsilon_{0}$ with $V$ depends on the way the electric field is induced in the system and on the characteristic of the impurity. To minimize the number of parameters in the model we present results with constant, $V$ independent, $\varepsilon_{0}$.
Having in mind the one impurity problem results, summarized in Fig. \ref{fig2}, the more relevant case of many impurities can be easily interpreted.

\section{Numerical Results for the Many Impurities Case}
Here we present results for the case of a small concentration of adatoms on the top layer. Calculations using Density Functional Theory show that, in the case of fluorine atoms, the adsorption energies on the two non-equivalent sites, $A_{1}$ and $B_{1}$, are almost equal. Some estimations, however, suggest that there could be a tiny energy gain for adatoms on the $B_{1}$ sites \cite{GySPrivComm}.
Interestingly, in other carbon based systems like monolayer graphene with substitutional nitrogen impurities, it has been observed that as the nitrogen concentration increases, impurities tend to be absorbed preferentially in one of the two equivalent  sublattices \cite{Zabet2014}. These self-organized structures of the nitrogen doped graphene are stabilized by the impurity-impurity interaction that favors impurities on the same sublattice, an effect that scales quickly with the impurity concentration \cite{Lawlor2014}.  
For the case of diluted fluorine adatoms on BLG there are no evidences of clustering on one sublattice. Moreover, the interaction between impurities on graphene is known to depend crucially on the type of impurity and on the adsorption geometry \cite{Gorman2013}. Based on these facts, in what follows we consider different impurity distributions, going from $50\%$ of the impurities in each sublattice to $100\%$ of them on the $B_{1}$ sites.  
We start with a detailed analysis of the LDOS at the impurity and at the different sites of the BLG.  
\subsection{Spectral densities}
To calculate $\rho_{\mathrm{imp}}(\omega)$ in the many impurities case we first use the Chebyshev polynomials method which has proven to be very efficient to deal with realistic impurity concentrations \cite{Weisse2006,Covaci2010,Yuan2010,Usaj2014}. The average impurity spectral density is then given by
\begin{equation}
\rho_{\mathrm{imp}}(\omega)=-\frac{1}{\pi}\langle\mathrm{Im}{\mathcal{G}_{ll}}\rangle_\mathrm{avg}\,,
\end{equation}
where $\langle\dots\rangle_\mathrm{avg}$ indicates the configurational average over the impurities. 

\begin{figure}[tb]
\includegraphics[width=0.95\columnwidth]{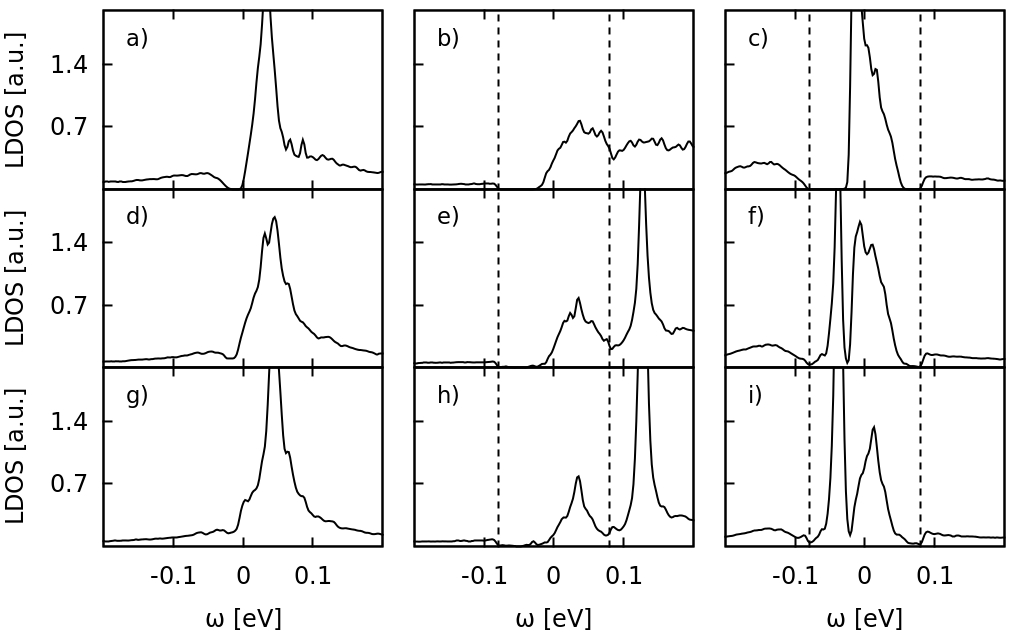}
\caption{Impurity spectral densities for different bias voltages and different impurity distributions. Top [(a), (b), (c)], central [(d), (e), (f)] and bottom [(g), (h), (i)] panels correspond to $100$\%, $75\%$  and $50\%$ of the impurities on the $B_{1}$ sublattice respectively.  Left [(a), (d), (g)], central [(b), (e), (h)] and right [(c), (f), (i)] columns correspond to $V=0$, $V=0.1$ eV and $V= -0.1$eV,  respectively.   Vertical dashed lines indicate the gap corresponding to the pristine BLG.}
\label{fig3}
\end{figure}

Figure \ref{fig3} shows $\rho_{\mathrm{imp}}(\omega)$ for a system with an impurity concentration $n_i=1/1800$ in a cluster with $8000$ impurities, different values of the parameter $V$ and different percentage of impurities on each sublattice. 
We consider first the less realistic, but simpler, case where  all impurities are adsorbed on the $B_{1}$ sites, top panels of Fig. \ref{fig3}. For $V=0$, as in the one impurity case,  we obtain a broad peak in $\rho_{\mathrm{imp}}(\omega)$ located near the renormalized energy $\bar{\varepsilon}_0$.  A remarkable detail is the emergence of a small gap for $\omega<0$. This gap is reminiscent of the gap that occurs in graphene monolayers for a finite concentration of impurities lying on the same sublattice \cite{Pereira2008,Cheianov2010,Abanin2010,Santos2014}.    The effect is due to a global inversion symmetry breaking due to the different structure of the $A$ and $B$ sublattices. 
In the thermodynamic limit, disordered systems would not present real gaps but energy windows with exponentially small DOS and it would be more appropriate to talk about pseudo-gaps rather than about real gaps.

For a gated system with  positive $V=0.1$ eV the gap induced in the pristine BLG is partially filled by impurity states. Within this gap, the impurities generate a band that extends from the bottom of the conduction band towards the centre of the gap and is separated by a pseudo-gap from the valence band. Conversely, for $V=-0.1$ eV there is a narrower impurity band close to the centre of the  BLG gap separated by pseudo-gaps from the conduction and valence bands. These structures can be understood straightforwardly from the shape of the bound states of the one impurity case.

\begin{figure}[tb]
\includegraphics[width=0.95\columnwidth]{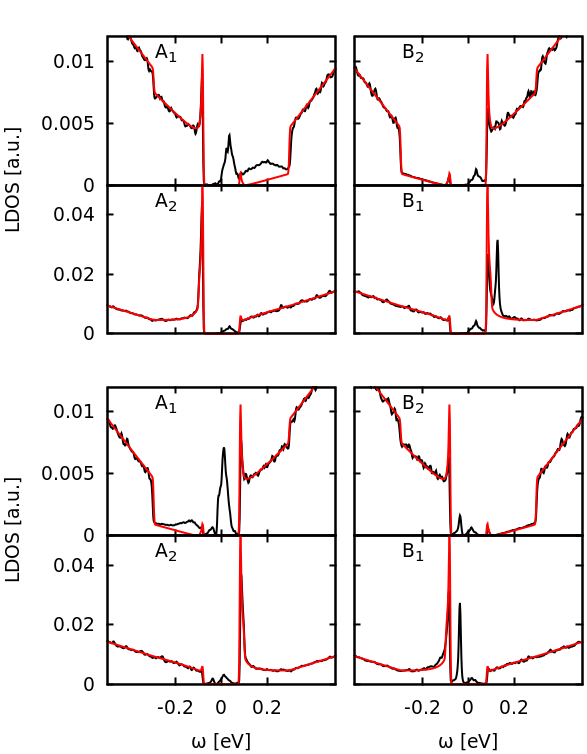}
\caption{Average LDOS of the four non equivalent C atoms (indicated in the figures) in a biased BLG with impurities distributed at $50$\%. The thin (red) lines are the corresponding LDOS of the pristine system. Upper and lower panels have $V=0.1$ eV and $V= -0.1$ eV, respectively. }
\label{fig4}
\end{figure}

The results are different if the impurities are distributed with the same probability on the two sublattices of the top layer, bottom panels of Fig. \ref{fig3}. For $V=0$ there is no pseudo-gap on top of the valence band. For positive $V$ an impurity band is formed within the BLG gap and a narrow resonance appears close to the bottom of the conduction band. The former is due to the impurities adsorbed on the $B_{1}$ sites while the later is due to the narrow resonance of the impurities on the $A_{1}$ sites (see Fig. \ref{fig2}c). Interestingly, for negative $V$ two separated and narrow impurity bands are formed within the BLG gap. Again, these bands are due to the impurities adsorbed on different sublattices, the lower energy one is narrower and comes from impurities on the $A_{1}$ sublattice. 

Other impurity distributions, like the one shown at the central panel of Fig. \ref{fig3} where $2/3$ of the impurities are on the $B_{1}$ sublattice, can be viewed in a first approximation as an interpolation between the two previous cases where the spectral weight of the $A_{1}$ and $B_{1}$ impurities change according to their concentration.
 As we show below, in gated BLG,  the impurity bands formed within the gap of the pristine sample are bands of strongly localized states.

In order to better characterize the effect of impurities on the electronic structure of the system we evaluate the average LDOS on the four non-equivalent sites of the BLG, $\rho_{Ai}$ ($\rho_{Bi}$). The results are presented in Fig. \ref{fig4}. 

Notably, in some cases, for $\varepsilon_{0} >0$ and $V\ne 0$, the LDOS of the valence band of the host BLG is almost unaffected by the impurities. In particular, the narrow van Hove singularity of the $A_{2}$ sites is essentially insensitive to the presence of the adsorbate. This suggests that at least in the valence band no strong localization effects occur with this type of impurity doping. As we show in next section, in gated samples, clear evidence of strong localization are observed for states within the gap and close to the impurity resonances occurring in the conduction band.

\subsection{Localization and transport properties}

To estimate the localization length
$\xi (\omega )$ we evaluate the two-point correlation function $|\mathcal{G}_{ij}(\omega)|^2$, 
 where  ${\mathcal{G}}_{ij}(\omega)=\langle\langle f_{i}^{},f_{j}^{\dag }\rangle\rangle$ is the retarded propagator 
from the impurity orbital at site $i$  to the one at site $j$. In the localized regime this quantity decreases exponentially when the distance $R_{ij}$ between impurities increases.\cite{Evers2008} For large $R_{ij}$ ($R_{ij}\gtrsim\xi$), the configurational average of its logarithm is well described by the following expression \cite{Li1989}
\begin{equation}
\label{lnG}
\langle\ln{|\mathcal{G}_{ij}(\omega)|^2}\rangle_\mathrm{avg}=\beta-2R_{ij}/\xi (\omega)-\alpha \ln{R_{ij}}\,,
\end{equation}
where $\alpha$ and $\beta$ are fitting parameters. An estimation of the localization length $\xi(\omega)$ then requires the evaluation of the impurity propagator $\mathcal{G}_{ij}(\omega )$ at large distances $R_{ij}$. As discussed in Ref. [\onlinecite{Usaj2014}], the Chebyshev polynomials method becomes numerically inefficient to this end. However, for long distances and low energy, the propagators of the pristine BLG can be evaluates analytically using the continuous approximation.  Defining the impurity propagator matrix $\bm{\mathcal{G}}$ with matrix elements $\mathcal{G}_{ij}(\omega)$ the Dyson equation reads
\begin{equation}
\left[(\omega+\ci0^+-\varepsilon _{0})\bm{I}-\gamma^{2}\tilde{\bm{g}}\right]\bm{\mathcal{G}}=\bm{I}\,,
\label{Dyson}
\end{equation}
where $\bm{I}$ is the unit matrix and $\tilde{\bm{g}}$ is a matrix whose elements are
the propagators of pristine graphene, $g_{i,j}(\omega)$, between $C$ sites $i$ and $j$ having an impurity on top. 
The quantity $\tilde{t}_{ij}=\gamma^2g_{i,j}(\omega)$ represents an effective (frequency dependent) hopping between impurities. The BLG retarded propagators take the form
\begin{widetext}
\begin{eqnarray}
g_{A_{1}i,A_{1}j}({\bm{R}},\omega)&=&\frac{4\pi}{\Omega_{BZ}}\cos({\bm{K}\cdot \bm{R}})\left[\mu_{1}^{AA}K_{0}\left(ik_{1}R\right)-\mu_{2}^{AA}K_{0}\left(ik_{2}R\right)\right]\,,\\
g_{A_{1}i,B_{1}j}({\bm{R}},\omega)&=&-\frac{4\pi}{\Omega_{BZ}}\sin({\bm{K}\cdot \bm{R}}+\theta_{R})\left[\mu_{1}^{AB}K_{1}\left(ik_{1}R\right)-\mu_{2}^{AB}K_{1}\left(ik_{2}R\right)\right]\,,\\
g_{B_{1}i,B_{1}j}({\bm{R}},\omega)&=&\frac{4\pi}{\Omega_{BZ}}\cos({\bm{K}\cdot \bm{R}})\left[\mu_{1}^{BB}K_{0}\left(ik_{1}R\right)-\mu_{2}^{BB}K_{0}\left(ik_{2}R\right)\right].
\end{eqnarray}
\end{widetext}
Here $K_{\upsilon}(x)$ is the $\upsilon$ order modified Bessel function of the second kind, $\Omega_{BZ}$ is the area of the first Brillouin Zone and the coefficients are
\begin{equation}
\mu_{j}^{AA}=\left(\frac{\left(\omega-V\right)\left(\omega+V\right)^{2}}{v_{F}^{4}\left(k_{1}^{2}-k_{2}^{2}\right)}+\frac{k_{j}^{2}\left(V-\omega\right)}{v_{F}^{2}\left(k_{1}^{2}-k_{2}^{2}\right)}\right),
\end{equation}
\begin{equation}
\mu_{j}^{AB}=ik_{j}\left(\frac{\left(V+\omega\right)^{2}}{v_{F}^{3}\left(k_{1}^{2}-k_{2}^{2}\right)}-\frac{k_{j}^{2}}{v_{F}\left(k_{1}^{2}-k_{2}^{2}\right)}\right),
\end{equation}
\begin{equation}
\mu_{j}^{BB}=\left(\frac{\left(V+\omega\right)\left(\omega^{2}-V^{2}-t_{\perp}^{2}\right)}{v_{F}^{4}\left(k_{1}^{2}-k_{2}^{2}\right)}+\frac{k_{j}^{2}\left(V-\omega\right)}{v_{F}^{2}\left(k_{1}^{2}-k_{2}^{2}\right)}\right),
\end{equation}
with $j=1,2$, $v_{F}$ is the Fermi velocity,
\begin{equation}
k_{1,2}^{2}=\frac{V^{2}+\omega^{2}}{v_{F}^{2}}\pm\frac{1}{2v_{F}^{2}}\sqrt{16V^{2}\omega^{2}+4\left(\omega^{2}-V^{2}\right)t_{\perp}^{2}}.
\end{equation}
and $\theta_R$ is the polar angle of the direction
of $\bm{R}$ with respect to the $x$ axis chosen to be along the direction of $K$-$K'$. 
In addition, $g_{B_{1}j,A_{1}i}({\bm{R}},\omega)=g_{A_{1}i,B_{1}j}({-\bm{R}},\omega)$.

\begin{figure}[tb]
\includegraphics[width=0.95\columnwidth]{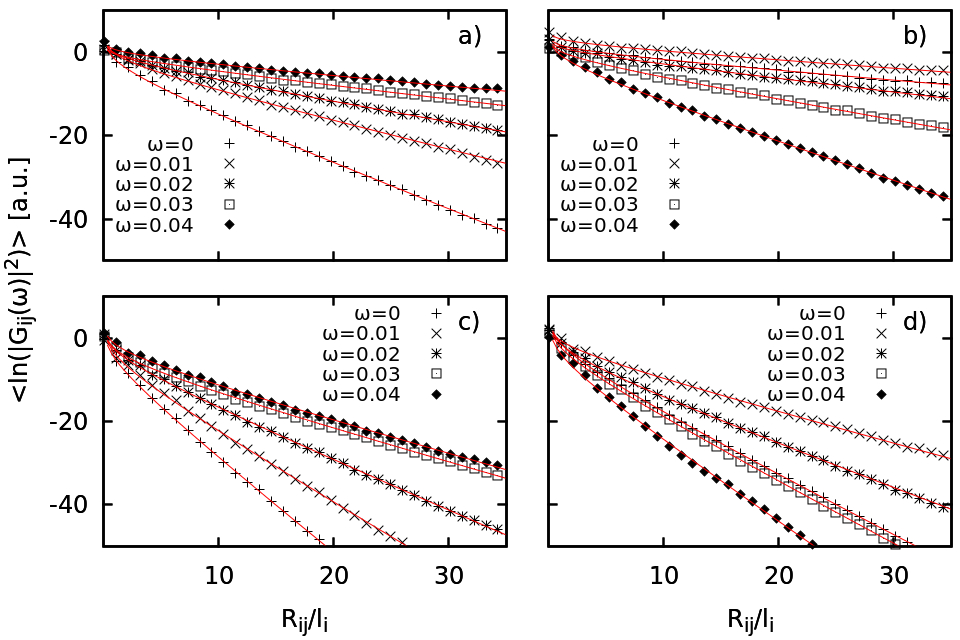}
\caption{Spatial dependence of $\langle\ln{|\mathcal{G}_{ij}(\omega)|^2}\rangle_\mathrm{avg}$ inside the gap of biased BLG. Top and bottom panels correspond to $100\%$ and $50\%$ of the impurities on the $B_{1}$ sublattice respectively. Left and right columns correspond to $V=0.1$ eV and $V=-0.1$ eV, respectively.}
\label{fig5}
\end{figure}

For a random distribution of impurities, we calculate the matrix $\tilde{\bm{g}}$ and obtain $ \bm{\mathcal{G}}$ from Eq. (\ref{Dyson}). We then take an average of $\ln{|\mathcal{G}_{ij}(\omega) |^2}$ for all sites $i$ and $j$ whose distance lies in a narrow window around a given value $R_{ij}$. In this procedure, to avoid finite size effects, we take site $i$ close to the centre on the cluster and neglect all sites $j$ lying close to the edges of the cluster. Finally, we make a configurational average by repeating the procedure with different impurity configurations. The obtained $\langle\ln{|\mathcal{G}_{ij}(\omega )|^2}\rangle_\mathrm{avg}$ versus $R_{ij}$ is then fitted using Eq. (\ref{lnG}) to obtain the localization length $\xi (\omega )$. Some of these fits are shown in Fig. \ref{fig5} for different values of the energy $\omega$ lying within gap of the biased BLG. 
In Fig. \ref{fig6} the localization length $\xi (\omega)$ for the gated and ungated BLG cases and different impurity distributions are shown for a cluster with typical radius of the order of $40 \ell_{i}$ where $\ell_{i}$ is the mean impurity-impurity distance. 
For the ungated system the localization length $\xi (\omega)$ presents a minimum in the conduction band for energies close to the renormalized energy $\bar{\varepsilon}_0$  of the impurity resonance. As the energy approaches the Dirac point from above, $\omega>0$, the localization length shows a fast increase exceeding the values for which our calculation gives reliable results (only localization lengths smaller than a fraction of the impurities cluster is considered). This behavior for $\omega>0$ is qualitative similar to what is observed in monolayer graphene.\cite{Usaj2014} In the valence band, there is a rapid increase of $\xi (\omega)$ as $|\omega|$ increases.

For the gated system, the impurity bands formed within the gap of the pristine BLG are strongly localized. In contrast, the states in the BLG bands tend to be much less localized, in particular in the valence band (consistent with the small sensitivity observed on the averaged LDOS, see Fig. \ref{fig4}) .

\subsection{Summary and discussion}

We have analyzed the effect of diluted adatoms on the electronic structure of gated and ungated bilayer graphene. The impurities are described as single orbital hybridized with the p$_{z}$ orbital of one of the C atoms of the top layer. We consider diluted systems, typically with impurity concentrations $n_{i}\approx 5\times10^{-4}$ and with different distributions on the two non-equivalent sites of the top graphene layer. 

\begin{figure}[tb]
\includegraphics[width=0.95\columnwidth]{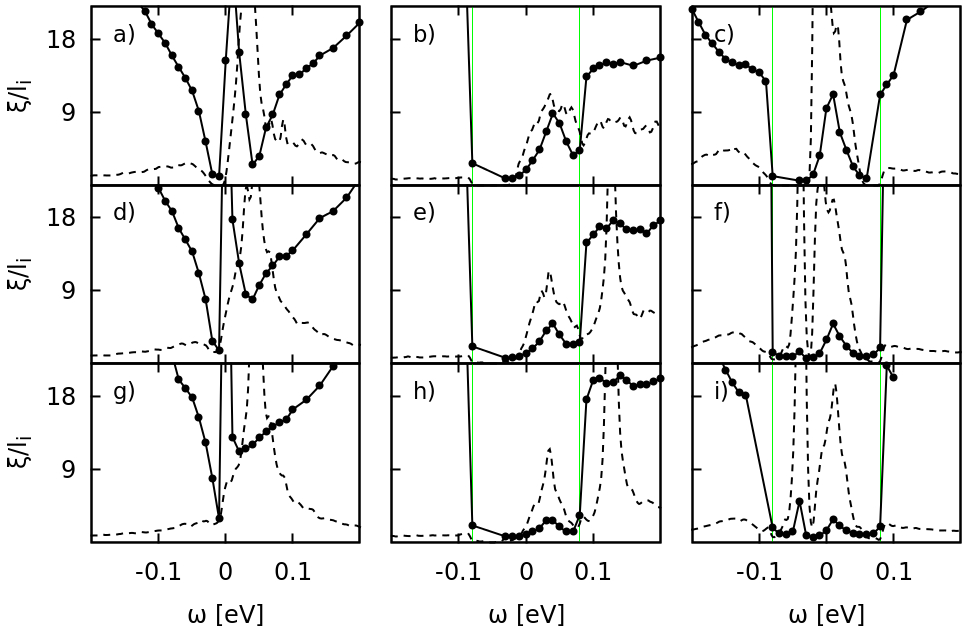}
\caption{Localization length in units of the average distance between impurities for bias voltages and different impurity distributions. Parameters like in Fig. \ref{fig3}. Dashed lines are the average impurity spectral densities.}
\label{fig6}
\end{figure}

In the diluted limit studied in this work, and due to the small adsorption energy difference of fluorine on the two different sites, the most probable impurity distributions would correspond to an almost random distribution of impurities on the two sublattices.  For the sake of concreteness, we consider the case of $50\%$ of the adatoms on each sublattice, illustrated in the bottom panels of Fig. \ref{fig6}, for our following concluding remarks.

The first observation is that for the same impurities and the same concentration, the localization length is larger in ungated BLG than in graphene. The localization length shows a minimum for energies close to the impurity resonance, there our results show that $\xi(\omega)$ is at least two times larger in BLG than in graphene. The behavior of $\xi(\omega)$ at the Dirac point $(\omega \approx 0)$ shows a structure with a sharp minimum. In second place, the effect of an electric field perpendicular to the sample depends on the polarity of the field. The field induces a gap in the pristine BLG that is partially filled by strongly localized impurity states. However, the structure, distribution and localization length of these states depend on the field polarization. For positive $V$ a single impurity band covers the upper part of the gap.  There, all states are strongly localized with a maximum of $\xi(\omega)$ at the centre of the band. The impurity spectral density shows a sharp resonance at the bottom of the conduction band. This resonance is due to localized states that are much more extended than those in the gap. In the valence band the localization length is too large for a good estimation with the system size used in the calculation.  For negative $V$ two narrow impurity bands, separated from each other and from the valence and conduction bands by small pseudo-gaps, are obtained. In both bands the localization length shows marked energy dependence with a maximum at the centre of each band. The states in the valence and conduction bands are much less localized. 

In systems with a weak energy dependence of the density of states and the localization length around the Fermi energy $\Ef$,   the resistance $\mathcal{R}(T)$ is expected to show the Mott's variable range hopping (VRH) behavior. In two dimensional system the VRH theory gives $\mathcal{R}(T)\propto \exp[(T_{0}/T)^{\frac{1}{3}}]$, where $T_{0}$ is a characteristic activation temperature given by 
\begin{equation}
T_{0}=\frac{\alpha}{k_\mathrm{B}\rho(\Ef)\xi ^{2}(\Ef)}\,.
\end{equation}
Here $\alpha$ is a numerical constant ($\alpha\approx 14$), $\rho(\Ef)$ and $\xi(\Ef)$ are the total density of states (DOS) and the localization length at the Fermi energy, respectively. In biased BLG, where two distinct strongly localized impurities bands may exist inside the gap, one could expect deviations of $\mathcal{R}(T)$ from a single VRH theory. In that case, a generalization of Eq.(15) to the case of two narrow impurity bands might be needed.

Finally, due to the dependence of the low energies electronic structure on the polarity of the electric field, the model predicts a dependence of the transport properties on the sign of $V$. Such asymmetry is not clearly observed in experiments with fluorinated graphene \cite{JunPrivComm}. If all impurities where adsorbed on the $B_{1}$ sublattice these asymmetries would be difficult to observe due to the similarities on the DOS and the localization length obtained for the two polarities, see top panels of Fig. \ref{fig6}. This scenario, however, is very unlikely. On the other hand, it has been reported that bilayer graphene samples grown on SiO$_{2}$/Si may show charge inhomogeneities with variations of the electronic density up to $10^{11}$cm$^{2}$. Such inhomogeneities, that locally shift the (electro) chemical potential in different regions of the sample, could also mask the asymmetries.

\section{Acknowledgements}
We acknowledge useful discussions with J. Sofo , J. Zhu and R. M. Guzm\'an Arellano. We thank financial support from PICT
Bicentenario 2010-1060 from ANPCyT, PIP 11220080101821 and 11220110100832 from CONICET and 06/C400 and 06/C415 SeCyT-UNC. HPOC
and GU acknowledge support from the ICTP. GU also acknowledges support from the Simons Foundation.



\begin{thebibliography}{46}%
\makeatletter
\providecommand \@ifxundefined [1]{%
 \@ifx{#1\undefined}
}%
\providecommand \@ifnum [1]{%
 \ifnum #1\expandafter \@firstoftwo
 \else \expandafter \@secondoftwo
 \fi
}%
\providecommand \@ifx [1]{%
 \ifx #1\expandafter \@firstoftwo
 \else \expandafter \@secondoftwo
 \fi
}%
\providecommand \natexlab [1]{#1}%
\providecommand \enquote  [1]{``#1''}%
\providecommand \bibnamefont  [1]{#1}%
\providecommand \bibfnamefont [1]{#1}%
\providecommand \citenamefont [1]{#1}%
\providecommand \href@noop [0]{\@secondoftwo}%
\providecommand \href [0]{\begingroup \@sanitize@url \@href}%
\providecommand \@href[1]{\@@startlink{#1}\@@href}%
\providecommand \@@href[1]{\endgroup#1\@@endlink}%
\providecommand \@sanitize@url [0]{\catcode `\\12\catcode `\$12\catcode
  `\&12\catcode `\#12\catcode `\^12\catcode `\_12\catcode `\%12\relax}%
\providecommand \@@startlink[1]{}%
\providecommand \@@endlink[0]{}%
\providecommand \url  [0]{\begingroup\@sanitize@url \@url }%
\providecommand \@url [1]{\endgroup\@href {#1}{\urlprefix }}%
\providecommand \urlprefix  [0]{URL }%
\providecommand \Eprint [0]{\href }%
\providecommand \doibase [0]{http://dx.doi.org/}%
\providecommand \selectlanguage [0]{\@gobble}%
\providecommand \bibinfo  [0]{\@secondoftwo}%
\providecommand \bibfield  [0]{\@secondoftwo}%
\providecommand \translation [1]{[#1]}%
\providecommand \BibitemOpen [0]{}%
\providecommand \bibitemStop [0]{}%
\providecommand \bibitemNoStop [0]{.\EOS\space}%
\providecommand \EOS [0]{\spacefactor3000\relax}%
\providecommand \BibitemShut  [1]{\csname bibitem#1\endcsname}%
\let\auto@bib@innerbib\@empty
\bibitem [{\citenamefont {{Castro Neto}}\ \emph {et~al.}(2009)\citenamefont
  {{Castro Neto}}, \citenamefont {Guinea}, \citenamefont {Peres}, \citenamefont
  {Novoselov},\ and\ \citenamefont {Geim}}]{CastroNeto-review}%
  \BibitemOpen
  \bibfield  {author} {\bibinfo {author} {\bibfnamefont {A.~H.}\ \bibnamefont
  {{Castro Neto}}}, \bibinfo {author} {\bibfnamefont {F.}~\bibnamefont
  {Guinea}}, \bibinfo {author} {\bibfnamefont {N.~M.~R.}\ \bibnamefont
  {Peres}}, \bibinfo {author} {\bibfnamefont {K.~S.}\ \bibnamefont
  {Novoselov}}, \ and\ \bibinfo {author} {\bibfnamefont {A.~K.}\ \bibnamefont
  {Geim}},\ }\href {\doibase 10.1103/RevModPhys.81.109} {\bibfield  {journal}
  {\bibinfo  {journal} {Rev. Mod. Phys.}\ }\textbf {\bibinfo {volume} {81}},\
  \bibinfo {eid} {109} (\bibinfo {year} {2009})},\ \bibinfo {note} {and refs.
  therein}\BibitemShut {NoStop}%
\bibitem [{\citenamefont {Das~Sarma}\ \emph {et~al.}(2011)\citenamefont
  {Das~Sarma}, \citenamefont {Adam}, \citenamefont {Hwang},\ and\ \citenamefont
  {Rossi}}]{DasSarma2011}%
  \BibitemOpen
  \bibfield  {author} {\bibinfo {author} {\bibfnamefont {S.}~\bibnamefont
  {Das~Sarma}}, \bibinfo {author} {\bibfnamefont {S.}~\bibnamefont {Adam}},
  \bibinfo {author} {\bibfnamefont {E.~H.}\ \bibnamefont {Hwang}}, \ and\
  \bibinfo {author} {\bibfnamefont {E.}~\bibnamefont {Rossi}},\ }\href
  {\doibase 10.1103/RevModPhys.83.407} {\bibfield  {journal} {\bibinfo
  {journal} {Rev. Mod. Phys.}\ }\textbf {\bibinfo {volume} {83}},\ \bibinfo
  {pages} {407} (\bibinfo {year} {2011})}\BibitemShut {NoStop}%
\bibitem [{\citenamefont {Beenakker}(2008)}]{Beenakker2008}%
  \BibitemOpen
  \bibfield  {author} {\bibinfo {author} {\bibfnamefont {C.~W.~J.}\
  \bibnamefont {Beenakker}},\ }\href {\doibase 10.1103/RevModPhys.80.1337}
  {\bibfield  {journal} {\bibinfo  {journal} {Rev. Mod. Phys.}\ }\textbf
  {\bibinfo {volume} {80}},\ \bibinfo {eid} {1337} (\bibinfo {year}
  {2008})}\BibitemShut {NoStop}%
\bibitem [{\citenamefont {Evers}\ and\ \citenamefont
  {Mirlin}(2008)}]{Evers2008}%
  \BibitemOpen
  \bibfield  {author} {\bibinfo {author} {\bibfnamefont {F.}~\bibnamefont
  {Evers}}\ and\ \bibinfo {author} {\bibfnamefont {A.}~\bibnamefont {Mirlin}},\
  }\href@noop {} {\bibfield  {journal} {\bibinfo  {journal} {Rev. Mod. Phys.}\
  }\textbf {\bibinfo {volume} {80}},\ \bibinfo {pages} {1355} (\bibinfo {year}
  {2008})}\BibitemShut {NoStop}%
\bibitem [{\citenamefont {Chan}\ \emph {et~al.}(2008)\citenamefont {Chan},
  \citenamefont {Neaton},\ and\ \citenamefont {Cohen}}]{Chan2008}%
  \BibitemOpen
  \bibfield  {author} {\bibinfo {author} {\bibfnamefont {K.~T.}\ \bibnamefont
  {Chan}}, \bibinfo {author} {\bibfnamefont {J.~B.}\ \bibnamefont {Neaton}}, \
  and\ \bibinfo {author} {\bibfnamefont {M.~L.}\ \bibnamefont {Cohen}},\ }\href
  {\doibase 10.1103/PhysRevB.77.235430} {\bibfield  {journal} {\bibinfo
  {journal} {Phys. Rev. B}\ }\textbf {\bibinfo {volume} {77}},\ \bibinfo
  {pages} {235430} (\bibinfo {year} {2008})}\BibitemShut {NoStop}%
\bibitem [{\citenamefont {Wehling}\ \emph {et~al.}(2009)\citenamefont
  {Wehling}, \citenamefont {Katsnelson},\ and\ \citenamefont
  {Lichtenstein}}]{Wehling2009}%
  \BibitemOpen
  \bibfield  {author} {\bibinfo {author} {\bibfnamefont {T.~O.}\ \bibnamefont
  {Wehling}}, \bibinfo {author} {\bibfnamefont {M.~I.}\ \bibnamefont
  {Katsnelson}}, \ and\ \bibinfo {author} {\bibfnamefont {A.~I.}\ \bibnamefont
  {Lichtenstein}},\ }\href {\doibase 10.1103/PhysRevB.80.085428} {\bibfield
  {journal} {\bibinfo  {journal} {Phys. Rev. B}\ }\textbf {\bibinfo {volume}
  {80}},\ \bibinfo {pages} {085428} (\bibinfo {year} {2009})}\BibitemShut
  {NoStop}%
\bibitem [{\citenamefont {Wehling}\ \emph
  {et~al.}(2010{\natexlab{a}})\citenamefont {Wehling}, \citenamefont
  {Balatsky}, \citenamefont {Katsnelson}, \citenamefont {Lichtenstein},\ and\
  \citenamefont {Rosch}}]{Wehling2010}%
  \BibitemOpen
  \bibfield  {author} {\bibinfo {author} {\bibfnamefont {T.~O.}\ \bibnamefont
  {Wehling}}, \bibinfo {author} {\bibfnamefont {A.~V.}\ \bibnamefont
  {Balatsky}}, \bibinfo {author} {\bibfnamefont {M.~I.}\ \bibnamefont
  {Katsnelson}}, \bibinfo {author} {\bibfnamefont {A.~I.}\ \bibnamefont
  {Lichtenstein}}, \ and\ \bibinfo {author} {\bibfnamefont {A.}~\bibnamefont
  {Rosch}},\ }\href {\doibase 10.1103/PhysRevB.81.115427} {\bibfield  {journal}
  {\bibinfo  {journal} {Phys. Rev. B}\ }\textbf {\bibinfo {volume} {81}},\
  \bibinfo {pages} {115427} (\bibinfo {year} {2010}{\natexlab{a}})}\BibitemShut
  {NoStop}%
\bibitem [{\citenamefont {Wehling}\ \emph
  {et~al.}(2010{\natexlab{b}})\citenamefont {Wehling}, \citenamefont {Yuan},
  \citenamefont {Lichtenstein}, \citenamefont {Geim},\ and\ \citenamefont
  {Katsnelson}}]{Wehling2010b}%
  \BibitemOpen
  \bibfield  {author} {\bibinfo {author} {\bibfnamefont {T.~O.}\ \bibnamefont
  {Wehling}}, \bibinfo {author} {\bibfnamefont {S.}~\bibnamefont {Yuan}},
  \bibinfo {author} {\bibfnamefont {A.~I.}\ \bibnamefont {Lichtenstein}},
  \bibinfo {author} {\bibfnamefont {A.~K.}\ \bibnamefont {Geim}}, \ and\
  \bibinfo {author} {\bibfnamefont {M.~I.}\ \bibnamefont {Katsnelson}},\
  }\href@noop {} {\bibfield  {journal} {\bibinfo  {journal} {Phys. Rev. Lett.}\
  }\textbf {\bibinfo {volume} {105}},\ \bibinfo {pages} {056802} (\bibinfo
  {year} {2010}{\natexlab{b}})}\BibitemShut {NoStop}%
\bibitem [{\citenamefont {Sofo}\ \emph {et~al.}(2012)\citenamefont {Sofo},
  \citenamefont {Usaj}, \citenamefont {Cornaglia}, \citenamefont {Suarez},
  \citenamefont {Hern{\'a}ndez-Nieves},\ and\ \citenamefont
  {Balseiro}}]{Sofo2012}%
  \BibitemOpen
  \bibfield  {author} {\bibinfo {author} {\bibfnamefont {J.}~\bibnamefont
  {Sofo}}, \bibinfo {author} {\bibfnamefont {G.}~\bibnamefont {Usaj}}, \bibinfo
  {author} {\bibfnamefont {P.~S.}\ \bibnamefont {Cornaglia}}, \bibinfo {author}
  {\bibfnamefont {A.}~\bibnamefont {Suarez}}, \bibinfo {author} {\bibfnamefont
  {A.~D.}\ \bibnamefont {Hern{\'a}ndez-Nieves}}, \ and\ \bibinfo {author}
  {\bibfnamefont {C.~A.}\ \bibnamefont {Balseiro}},\ }\href@noop {} {\bibfield
  {journal} {\bibinfo  {journal} {Phys. Rev. B}\ }\textbf {\bibinfo {volume}
  {85}},\ \bibinfo {pages} {115405} (\bibinfo {year} {2012})}\BibitemShut
  {NoStop}%
\bibitem [{\citenamefont {Roche}\ \emph {et~al.}(2012)\citenamefont {Roche},
  \citenamefont {Leconte}, \citenamefont {Ortmann}, \citenamefont {Lherbier},
  \citenamefont {Soriano},\ and\ \citenamefont {Charlier}}]{Roche2012}%
  \BibitemOpen
  \bibfield  {author} {\bibinfo {author} {\bibfnamefont {S.}~\bibnamefont
  {Roche}}, \bibinfo {author} {\bibfnamefont {N.}~\bibnamefont {Leconte}},
  \bibinfo {author} {\bibfnamefont {F.}~\bibnamefont {Ortmann}}, \bibinfo
  {author} {\bibfnamefont {A.}~\bibnamefont {Lherbier}}, \bibinfo {author}
  {\bibfnamefont {D.}~\bibnamefont {Soriano}}, \ and\ \bibinfo {author}
  {\bibfnamefont {J.-C.}\ \bibnamefont {Charlier}},\ }\href {\doibase
  http://dx.doi.org/10.1016/j.ssc.2012.04.030} {\bibfield  {journal} {\bibinfo
  {journal} {Solid State Comm.}\ }\textbf {\bibinfo {volume} {152}},\ \bibinfo
  {pages} {1404 } (\bibinfo {year} {2012})}\BibitemShut {NoStop}%
\bibitem [{\citenamefont {Matis}\ \emph {et~al.}(2012)\citenamefont {Matis},
  \citenamefont {Bulat}, \citenamefont {Friedman}, \citenamefont {Houston},\
  and\ \citenamefont {Baldwin}}]{Matis2012}%
  \BibitemOpen
  \bibfield  {author} {\bibinfo {author} {\bibfnamefont {B.}~\bibnamefont
  {Matis}}, \bibinfo {author} {\bibfnamefont {F.}~\bibnamefont {Bulat}},
  \bibinfo {author} {\bibfnamefont {A.}~\bibnamefont {Friedman}}, \bibinfo
  {author} {\bibfnamefont {B.}~\bibnamefont {Houston}}, \ and\ \bibinfo
  {author} {\bibfnamefont {J.}~\bibnamefont {Baldwin}},\ }\href@noop {}
  {\bibfield  {journal} {\bibinfo  {journal} {Phys. Rev. B}\ }\textbf {\bibinfo
  {volume} {85}} (\bibinfo {year} {2012})}\BibitemShut {NoStop}%
\bibitem [{\citenamefont {Guillemette}\ \emph {et~al.}(2013)\citenamefont
  {Guillemette}, \citenamefont {Sabri}, \citenamefont {Wu}, \citenamefont
  {Bennaceur}, \citenamefont {Gaskell}, \citenamefont {Savard}, \citenamefont
  {L{\'e}vesque}, \citenamefont {Mahvash}, \citenamefont {Guermoune},
  \citenamefont {Siaj}, \citenamefont {Martel}, \citenamefont {Szkopek},\ and\
  \citenamefont {Gervais}}]{Guillemette2013}%
  \BibitemOpen
  \bibfield  {author} {\bibinfo {author} {\bibfnamefont {J.}~\bibnamefont
  {Guillemette}}, \bibinfo {author} {\bibfnamefont {S.~S.}\ \bibnamefont
  {Sabri}}, \bibinfo {author} {\bibfnamefont {B.}~\bibnamefont {Wu}}, \bibinfo
  {author} {\bibfnamefont {K.}~\bibnamefont {Bennaceur}}, \bibinfo {author}
  {\bibfnamefont {P.~E.}\ \bibnamefont {Gaskell}}, \bibinfo {author}
  {\bibfnamefont {M.}~\bibnamefont {Savard}}, \bibinfo {author} {\bibfnamefont
  {P.~L.}\ \bibnamefont {L{\'e}vesque}}, \bibinfo {author} {\bibfnamefont
  {F.}~\bibnamefont {Mahvash}}, \bibinfo {author} {\bibfnamefont
  {A.}~\bibnamefont {Guermoune}}, \bibinfo {author} {\bibfnamefont
  {M.}~\bibnamefont {Siaj}}, \bibinfo {author} {\bibfnamefont {R.}~\bibnamefont
  {Martel}}, \bibinfo {author} {\bibfnamefont {T.}~\bibnamefont {Szkopek}}, \
  and\ \bibinfo {author} {\bibfnamefont {G.}~\bibnamefont {Gervais}},\
  }\href@noop {} {\bibfield  {journal} {\bibinfo  {journal} {Phys. Rev. Lett.}\
  }\textbf {\bibinfo {volume} {110}},\ \bibinfo {pages} {176801} (\bibinfo
  {year} {2013})}\BibitemShut {NoStop}%
\bibitem [{\citenamefont {Hong}\ \emph {et~al.}(2011)\citenamefont {Hong},
  \citenamefont {Cheng}, \citenamefont {Herding},\ and\ \citenamefont
  {Zhu}}]{Hong2011}%
  \BibitemOpen
  \bibfield  {author} {\bibinfo {author} {\bibfnamefont {X.}~\bibnamefont
  {Hong}}, \bibinfo {author} {\bibfnamefont {S.~H.}\ \bibnamefont {Cheng}},
  \bibinfo {author} {\bibfnamefont {C.}~\bibnamefont {Herding}}, \ and\
  \bibinfo {author} {\bibfnamefont {J.}~\bibnamefont {Zhu}},\ }\href@noop {}
  {\bibfield  {journal} {\bibinfo  {journal} {Phys. Rev. B}\ }\textbf {\bibinfo
  {volume} {83}},\ \bibinfo {pages} {085410} (\bibinfo {year}
  {2011})}\BibitemShut {NoStop}%
\bibitem [{\citenamefont {Sofo}\ \emph {et~al.}(2011)\citenamefont {Sofo},
  \citenamefont {Suarez}, \citenamefont {Usaj}, \citenamefont {Cornaglia},
  \citenamefont {Hern\'andez-Nieves},\ and\ \citenamefont
  {Balseiro}}]{Sofo2011}%
  \BibitemOpen
  \bibfield  {author} {\bibinfo {author} {\bibfnamefont {J.~O.}\ \bibnamefont
  {Sofo}}, \bibinfo {author} {\bibfnamefont {A.~M.}\ \bibnamefont {Suarez}},
  \bibinfo {author} {\bibfnamefont {G.}~\bibnamefont {Usaj}}, \bibinfo {author}
  {\bibfnamefont {P.~S.}\ \bibnamefont {Cornaglia}}, \bibinfo {author}
  {\bibfnamefont {A.~D.}\ \bibnamefont {Hern\'andez-Nieves}}, \ and\ \bibinfo
  {author} {\bibfnamefont {C.~A.}\ \bibnamefont {Balseiro}},\ }\href {\doibase
  10.1103/PhysRevB.83.081411} {\bibfield  {journal} {\bibinfo  {journal} {Phys.
  Rev. B}\ }\textbf {\bibinfo {volume} {83}},\ \bibinfo {pages} {081411}
  (\bibinfo {year} {2011})}\BibitemShut {NoStop}%
\bibitem [{\citenamefont {Chan}\ \emph {et~al.}(2011)\citenamefont {Chan},
  \citenamefont {Lee},\ and\ \citenamefont {Cohen}}]{Chan2011}%
  \BibitemOpen
  \bibfield  {author} {\bibinfo {author} {\bibfnamefont {K.~T.}\ \bibnamefont
  {Chan}}, \bibinfo {author} {\bibfnamefont {H.}~\bibnamefont {Lee}}, \ and\
  \bibinfo {author} {\bibfnamefont {M.~L.}\ \bibnamefont {Cohen}},\ }\href
  {\doibase 10.1103/PhysRevB.84.165419} {\bibfield  {journal} {\bibinfo
  {journal} {Phys. Rev. B}\ }\textbf {\bibinfo {volume} {84}},\ \bibinfo
  {pages} {165419} (\bibinfo {year} {2011})}\BibitemShut {NoStop}%
\bibitem [{\citenamefont {Guzm{\'a}n-Arellano}\ \emph
  {et~al.}(2014)\citenamefont {Guzm{\'a}n-Arellano}, \citenamefont
  {Hern{\'a}ndez-Nieves}, \citenamefont {Balseiro},\ and\ \citenamefont
  {Usaj}}]{Guzman2014}%
  \BibitemOpen
  \bibfield  {author} {\bibinfo {author} {\bibfnamefont {R.~M.}\ \bibnamefont
  {Guzm{\'a}n-Arellano}}, \bibinfo {author} {\bibfnamefont {A.~D.}\
  \bibnamefont {Hern{\'a}ndez-Nieves}}, \bibinfo {author} {\bibfnamefont
  {C.~A.}\ \bibnamefont {Balseiro}}, \ and\ \bibinfo {author} {\bibfnamefont
  {G.}~\bibnamefont {Usaj}},\ }\href {\doibase
  http://dx.doi.org/10.1063/1.4896511} {\bibfield  {journal} {\bibinfo
  {journal} {Appl. Phys. Lett.}\ }\textbf {\bibinfo {volume} {105}},\ \bibinfo
  {eid} {121606} (\bibinfo {year} {2014})}\BibitemShut {NoStop}%
\bibitem [{\citenamefont {Aleiner}\ and\ \citenamefont
  {Efetov}(2006)}]{Aleiner2006}%
  \BibitemOpen
  \bibfield  {author} {\bibinfo {author} {\bibfnamefont {I.~L.}\ \bibnamefont
  {Aleiner}}\ and\ \bibinfo {author} {\bibfnamefont {K.~B.}\ \bibnamefont
  {Efetov}},\ }\href@noop {} {\bibfield  {journal} {\bibinfo  {journal} {Phys.
  Rev. Lett.}\ }\textbf {\bibinfo {volume} {97}},\ \bibinfo {pages} {236801}
  (\bibinfo {year} {2006})}\BibitemShut {NoStop}%
\bibitem [{\citenamefont {Ostrovsky}\ \emph {et~al.}(2006)\citenamefont
  {Ostrovsky}, \citenamefont {Gornyi},\ and\ \citenamefont
  {Mirlin}}]{Ostrovsky2006}%
  \BibitemOpen
  \bibfield  {author} {\bibinfo {author} {\bibfnamefont {P.~M.}\ \bibnamefont
  {Ostrovsky}}, \bibinfo {author} {\bibfnamefont {I.~V.}\ \bibnamefont
  {Gornyi}}, \ and\ \bibinfo {author} {\bibfnamefont {A.~D.}\ \bibnamefont
  {Mirlin}},\ }\href {\doibase 10.1103/PhysRevB.74.235443} {\bibfield
  {journal} {\bibinfo  {journal} {Phys. Rev. B}\ }\textbf {\bibinfo {volume}
  {74}},\ \bibinfo {pages} {235443} (\bibinfo {year} {2006})}\BibitemShut
  {NoStop}%
\bibitem [{\citenamefont {Ostrovsky}\ \emph {et~al.}(2007)\citenamefont
  {Ostrovsky}, \citenamefont {Gornyi},\ and\ \citenamefont
  {Mirlin}}]{Ostrovsky2007}%
  \BibitemOpen
  \bibfield  {author} {\bibinfo {author} {\bibfnamefont {P.~M.}\ \bibnamefont
  {Ostrovsky}}, \bibinfo {author} {\bibfnamefont {I.~V.}\ \bibnamefont
  {Gornyi}}, \ and\ \bibinfo {author} {\bibfnamefont {A.~D.}\ \bibnamefont
  {Mirlin}},\ }\href@noop {} {\bibfield  {journal} {\bibinfo  {journal} {Eur.
  Phys. J. Spec. Top.}\ }\textbf {\bibinfo {volume} {148}},\ \bibinfo {pages}
  {63} (\bibinfo {year} {2007})}\BibitemShut {NoStop}%
\bibitem [{\citenamefont {Mirlin}\ \emph {et~al.}(2010)\citenamefont {Mirlin},
  \citenamefont {Evers}, \citenamefont {Gornyi},\ and\ \citenamefont
  {Ostrovsky}}]{Mirlin2010}%
  \BibitemOpen
  \bibfield  {author} {\bibinfo {author} {\bibfnamefont {A.~D.}\ \bibnamefont
  {Mirlin}}, \bibinfo {author} {\bibfnamefont {F.}~\bibnamefont {Evers}},
  \bibinfo {author} {\bibfnamefont {I.~V.}\ \bibnamefont {Gornyi}}, \ and\
  \bibinfo {author} {\bibfnamefont {P.~M.}\ \bibnamefont {Ostrovsky}},\
  }\href@noop {} {\bibfield  {journal} {\bibinfo  {journal} {Int. J. of Mod.
  Phys. B}\ }\textbf {\bibinfo {volume} {24}},\ \bibinfo {pages} {1577}
  (\bibinfo {year} {2010})}\BibitemShut {NoStop}%
\bibitem [{\citenamefont {K{\"o}nig}\ \emph {et~al.}(2012)\citenamefont
  {K{\"o}nig}, \citenamefont {Ostrovsky}, \citenamefont {Protopopov},\ and\
  \citenamefont {Mirlin}}]{Konig2012}%
  \BibitemOpen
  \bibfield  {author} {\bibinfo {author} {\bibfnamefont {E.~J.}\ \bibnamefont
  {K{\"o}nig}}, \bibinfo {author} {\bibfnamefont {P.~M.}\ \bibnamefont
  {Ostrovsky}}, \bibinfo {author} {\bibfnamefont {I.~V.}\ \bibnamefont
  {Protopopov}}, \ and\ \bibinfo {author} {\bibfnamefont {A.~D.}\ \bibnamefont
  {Mirlin}},\ }\href@noop {} {\bibfield  {journal} {\bibinfo  {journal} {Phys.
  Rev. B}\ }\textbf {\bibinfo {volume} {85}},\ \bibinfo {pages} {195130}
  (\bibinfo {year} {2012})}\BibitemShut {NoStop}%
\bibitem [{\citenamefont {Gattenloehner}\ \emph {et~al.}(2013)\citenamefont
  {Gattenloehner}, \citenamefont {Hannes}, \citenamefont {Ostrovsky},
  \citenamefont {Gornyi}, \citenamefont {Mirlin},\ and\ \citenamefont
  {Titov}}]{Gattenloehner2013}%
  \BibitemOpen
  \bibfield  {author} {\bibinfo {author} {\bibfnamefont {S.}~\bibnamefont
  {Gattenloehner}}, \bibinfo {author} {\bibfnamefont {W.~R.}\ \bibnamefont
  {Hannes}}, \bibinfo {author} {\bibfnamefont {P.~M.}\ \bibnamefont
  {Ostrovsky}}, \bibinfo {author} {\bibfnamefont {I.~V.}\ \bibnamefont
  {Gornyi}}, \bibinfo {author} {\bibfnamefont {A.~D.}\ \bibnamefont {Mirlin}},
  \ and\ \bibinfo {author} {\bibfnamefont {M.}~\bibnamefont {Titov}},\
  }\href@noop {} {\bibfield  {journal} {\bibinfo  {journal} {arXiv.org}\ }
  (\bibinfo {year} {2013})},\ \Eprint {http://arxiv.org/abs/1306.5686v1}
  {1306.5686v1} \BibitemShut {NoStop}%
\bibitem [{\citenamefont {Cresti}\ \emph {et~al.}(2013)\citenamefont {Cresti},
  \citenamefont {Ortmann}, \citenamefont {Louvet}, \citenamefont {Van~Tuan},\
  and\ \citenamefont {Roche}}]{Cresti2013}%
  \BibitemOpen
  \bibfield  {author} {\bibinfo {author} {\bibfnamefont {A.}~\bibnamefont
  {Cresti}}, \bibinfo {author} {\bibfnamefont {F.}~\bibnamefont {Ortmann}},
  \bibinfo {author} {\bibfnamefont {T.}~\bibnamefont {Louvet}}, \bibinfo
  {author} {\bibfnamefont {D.}~\bibnamefont {Van~Tuan}}, \ and\ \bibinfo
  {author} {\bibfnamefont {S.}~\bibnamefont {Roche}},\ }\href@noop {}
  {\bibfield  {journal} {\bibinfo  {journal} {Phys. Rev. Lett.}\ }\textbf
  {\bibinfo {volume} {110}},\ \bibinfo {pages} {196601} (\bibinfo {year}
  {2013})}\BibitemShut {NoStop}%
\bibitem [{\citenamefont {Usaj}\ \emph {et~al.}(2014)\citenamefont {Usaj},
  \citenamefont {Cornaglia},\ and\ \citenamefont {Balseiro}}]{Usaj2014}%
  \BibitemOpen
  \bibfield  {author} {\bibinfo {author} {\bibfnamefont {G.}~\bibnamefont
  {Usaj}}, \bibinfo {author} {\bibfnamefont {P.~S.}\ \bibnamefont {Cornaglia}},
  \ and\ \bibinfo {author} {\bibfnamefont {C.~A.}\ \bibnamefont {Balseiro}},\
  }\href {\doibase 10.1103/PhysRevB.89.085405} {\bibfield  {journal} {\bibinfo
  {journal} {Phys. Rev. B}\ }\textbf {\bibinfo {volume} {89}},\ \bibinfo
  {pages} {085405} (\bibinfo {year} {2014})}\BibitemShut {NoStop}%
\bibitem [{\citenamefont {Nilsson}\ \emph {et~al.}(2008)\citenamefont
  {Nilsson}, \citenamefont {Castro~Neto}, \citenamefont {Guinea},\ and\
  \citenamefont {Peres}}]{Nilsson2008}%
  \BibitemOpen
  \bibfield  {author} {\bibinfo {author} {\bibfnamefont {J.}~\bibnamefont
  {Nilsson}}, \bibinfo {author} {\bibfnamefont {A.~H.}\ \bibnamefont
  {Castro~Neto}}, \bibinfo {author} {\bibfnamefont {F.}~\bibnamefont {Guinea}},
  \ and\ \bibinfo {author} {\bibfnamefont {N.~M.~R.}\ \bibnamefont {Peres}},\
  }\href {\doibase 10.1103/PhysRevB.78.045405} {\bibfield  {journal} {\bibinfo
  {journal} {Phys. Rev. B}\ }\textbf {\bibinfo {volume} {78}},\ \bibinfo
  {pages} {045405} (\bibinfo {year} {2008})}\BibitemShut {NoStop}%
\bibitem [{\citenamefont {Castro}\ \emph {et~al.}(2010)\citenamefont {Castro},
  \citenamefont {Novoselov}, \citenamefont {Morozov}, \citenamefont {Peres},
  \citenamefont {dos Santos}, \citenamefont {Nilsson}, \citenamefont {Guinea},
  \citenamefont {Geim},\ and\ \citenamefont {Neto}}]{Castro2010}%
  \BibitemOpen
  \bibfield  {author} {\bibinfo {author} {\bibfnamefont {E.~V.}\ \bibnamefont
  {Castro}}, \bibinfo {author} {\bibfnamefont {K.~S.}\ \bibnamefont
  {Novoselov}}, \bibinfo {author} {\bibfnamefont {S.~V.}\ \bibnamefont
  {Morozov}}, \bibinfo {author} {\bibfnamefont {N.~M.~R.}\ \bibnamefont
  {Peres}}, \bibinfo {author} {\bibfnamefont {J.~M. B.~L.}\ \bibnamefont {dos
  Santos}}, \bibinfo {author} {\bibfnamefont {J.}~\bibnamefont {Nilsson}},
  \bibinfo {author} {\bibfnamefont {F.}~\bibnamefont {Guinea}}, \bibinfo
  {author} {\bibfnamefont {A.~K.}\ \bibnamefont {Geim}}, \ and\ \bibinfo
  {author} {\bibfnamefont {A.~H.~C.}\ \bibnamefont {Neto}},\ }\href
  {http://stacks.iop.org/0953-8984/22/i=17/a=175503} {\bibfield  {journal}
  {\bibinfo  {journal} {Journal of Physics: Condensed Matter}\ }\textbf
  {\bibinfo {volume} {22}},\ \bibinfo {pages} {175503} (\bibinfo {year}
  {2010})}\BibitemShut {NoStop}%
\bibitem [{\citenamefont {McCann}\ and\ \citenamefont
  {Koshino}(2013)}]{McCann2013}%
  \BibitemOpen
  \bibfield  {author} {\bibinfo {author} {\bibfnamefont {E.}~\bibnamefont
  {McCann}}\ and\ \bibinfo {author} {\bibfnamefont {M.}~\bibnamefont
  {Koshino}},\ }\href {http://stacks.iop.org/0034-4885/76/i=5/a=056503}
  {\bibfield  {journal} {\bibinfo  {journal} {Reports on Progress in Physics}\
  }\textbf {\bibinfo {volume} {76}},\ \bibinfo {pages} {056503} (\bibinfo
  {year} {2013})}\BibitemShut {NoStop}%
\bibitem [{\citenamefont {Castro}\ \emph {et~al.}(2007)\citenamefont {Castro},
  \citenamefont {Novoselov}, \citenamefont {Morozov}, \citenamefont {Peres},
  \citenamefont {dos Santos}, \citenamefont {Nilsson}, \citenamefont {Guinea},
  \citenamefont {Geim},\ and\ \citenamefont {Neto}}]{Castro2007}%
  \BibitemOpen
  \bibfield  {author} {\bibinfo {author} {\bibfnamefont {E.~V.}\ \bibnamefont
  {Castro}}, \bibinfo {author} {\bibfnamefont {K.~S.}\ \bibnamefont
  {Novoselov}}, \bibinfo {author} {\bibfnamefont {S.~V.}\ \bibnamefont
  {Morozov}}, \bibinfo {author} {\bibfnamefont {N.~M.~R.}\ \bibnamefont
  {Peres}}, \bibinfo {author} {\bibfnamefont {J.~M. B.~L.}\ \bibnamefont {dos
  Santos}}, \bibinfo {author} {\bibfnamefont {J.}~\bibnamefont {Nilsson}},
  \bibinfo {author} {\bibfnamefont {F.}~\bibnamefont {Guinea}}, \bibinfo
  {author} {\bibfnamefont {A.~K.}\ \bibnamefont {Geim}}, \ and\ \bibinfo
  {author} {\bibfnamefont {A.~H.~C.}\ \bibnamefont {Neto}},\ }\href {\doibase
  10.1103/PhysRevLett.99.216802} {\bibfield  {journal} {\bibinfo  {journal}
  {Phys. Rev. Lett.}\ }\textbf {\bibinfo {volume} {99}},\ \bibinfo {pages}
  {216802} (\bibinfo {year} {2007})}\BibitemShut {NoStop}%
\bibitem [{\citenamefont {McCann}(2006)}]{McCann2006}%
  \BibitemOpen
  \bibfield  {author} {\bibinfo {author} {\bibfnamefont {E.}~\bibnamefont
  {McCann}},\ }\href {\doibase 10.1103/PhysRevB.74.161403} {\bibfield
  {journal} {\bibinfo  {journal} {Phys. Rev. B}\ }\textbf {\bibinfo {volume}
  {74}},\ \bibinfo {pages} {161403} (\bibinfo {year} {2006})}\BibitemShut
  {NoStop}%
\bibitem [{\citenamefont {Min}\ \emph {et~al.}(2007)\citenamefont {Min},
  \citenamefont {Sahu}, \citenamefont {Banerjee},\ and\ \citenamefont
  {MacDonald}}]{Min2007}%
  \BibitemOpen
  \bibfield  {author} {\bibinfo {author} {\bibfnamefont {H.}~\bibnamefont
  {Min}}, \bibinfo {author} {\bibfnamefont {B.}~\bibnamefont {Sahu}}, \bibinfo
  {author} {\bibfnamefont {S.~K.}\ \bibnamefont {Banerjee}}, \ and\ \bibinfo
  {author} {\bibfnamefont {A.~H.}\ \bibnamefont {MacDonald}},\ }\href {\doibase
  10.1103/PhysRevB.75.155115} {\bibfield  {journal} {\bibinfo  {journal} {Phys.
  Rev. B}\ }\textbf {\bibinfo {volume} {75}},\ \bibinfo {pages} {155115}
  (\bibinfo {year} {2007})}\BibitemShut {NoStop}%
\bibitem [{\citenamefont {Taychatanapat}\ and\ \citenamefont
  {Jarillo-Herrero}(2010)}]{Taychatanapat2010}%
  \BibitemOpen
  \bibfield  {author} {\bibinfo {author} {\bibfnamefont {T.}~\bibnamefont
  {Taychatanapat}}\ and\ \bibinfo {author} {\bibfnamefont {P.}~\bibnamefont
  {Jarillo-Herrero}},\ }\href {\doibase 10.1103/PhysRevLett.105.166601}
  {\bibfield  {journal} {\bibinfo  {journal} {Phys. Rev. Lett.}\ }\textbf
  {\bibinfo {volume} {105}},\ \bibinfo {pages} {166601} (\bibinfo {year}
  {2010})}\BibitemShut {NoStop}%
\bibitem [{\citenamefont {Dahal}\ \emph {et~al.}(2008)\citenamefont {Dahal},
  \citenamefont {Balatsky},\ and\ \citenamefont {Zhu}}]{Dahal2008}%
  \BibitemOpen
  \bibfield  {author} {\bibinfo {author} {\bibfnamefont {H.~P.}\ \bibnamefont
  {Dahal}}, \bibinfo {author} {\bibfnamefont {A.~V.}\ \bibnamefont {Balatsky}},
  \ and\ \bibinfo {author} {\bibfnamefont {J.-X.}\ \bibnamefont {Zhu}},\ }\href
  {\doibase 10.1103/PhysRevB.77.115114} {\bibfield  {journal} {\bibinfo
  {journal} {Phys. Rev. B}\ }\textbf {\bibinfo {volume} {77}},\ \bibinfo
  {pages} {115114} (\bibinfo {year} {2008})}\BibitemShut {NoStop}%
\bibitem [{\citenamefont {Mkhitaryan}\ and\ \citenamefont
  {Mishchenko}(2013)}]{Mkhitaryan2013}%
  \BibitemOpen
  \bibfield  {author} {\bibinfo {author} {\bibfnamefont {V.~V.}\ \bibnamefont
  {Mkhitaryan}}\ and\ \bibinfo {author} {\bibfnamefont {E.~G.}\ \bibnamefont
  {Mishchenko}},\ }\href {\doibase 10.1103/PhysRevLett.110.086805} {\bibfield
  {journal} {\bibinfo  {journal} {Phys. Rev. Lett.}\ }\textbf {\bibinfo
  {volume} {110}},\ \bibinfo {pages} {086805} (\bibinfo {year}
  {2013})}\BibitemShut {NoStop}%
\bibitem [{\citenamefont {{Guzm\'an Arellano}}\ and\ \citenamefont
  {Sofo}(2014)}]{GySPrivComm}%
  \BibitemOpen
  \bibfield  {author} {\bibinfo {author} {\bibfnamefont {R.~M.}\ \bibnamefont
  {{Guzm\'an Arellano}}}\ and\ \bibinfo {author} {\bibfnamefont {J.~O.}\
  \bibnamefont {Sofo}},\ }\href@noop {} {}\bibinfo {howpublished} {private
  communications} (\bibinfo {year} {2014})\BibitemShut {NoStop}%
\bibitem [{\citenamefont {Zabet-Khosousi}\ \emph {et~al.}(2014)\citenamefont
  {Zabet-Khosousi}, \citenamefont {Zhao}, \citenamefont {Pálová},
  \citenamefont {Hybertsen}, \citenamefont {Reichman}, \citenamefont
  {Pasupathy},\ and\ \citenamefont {Flynn}}]{Zabet2014}%
  \BibitemOpen
  \bibfield  {author} {\bibinfo {author} {\bibfnamefont {A.}~\bibnamefont
  {Zabet-Khosousi}}, \bibinfo {author} {\bibfnamefont {L.}~\bibnamefont
  {Zhao}}, \bibinfo {author} {\bibfnamefont {L.}~\bibnamefont {Pálová}},
  \bibinfo {author} {\bibfnamefont {M.~S.}\ \bibnamefont {Hybertsen}}, \bibinfo
  {author} {\bibfnamefont {D.~R.}\ \bibnamefont {Reichman}}, \bibinfo {author}
  {\bibfnamefont {A.~N.}\ \bibnamefont {Pasupathy}}, \ and\ \bibinfo {author}
  {\bibfnamefont {G.~W.}\ \bibnamefont {Flynn}},\ }\href {\doibase
  10.1021/ja408463g} {\bibfield  {journal} {\bibinfo  {journal} {Journal of the
  American Chemical Society}\ }\textbf {\bibinfo {volume} {136}},\ \bibinfo
  {pages} {1391} (\bibinfo {year} {2014})}\BibitemShut {NoStop}%
\bibitem [{\citenamefont {Lawlor}\ \emph {et~al.}(2014)\citenamefont {Lawlor},
  \citenamefont {Gorman}, \citenamefont {Power}, \citenamefont {Bezerra},\ and\
  \citenamefont {Ferreira}}]{Lawlor2014}%
  \BibitemOpen
  \bibfield  {author} {\bibinfo {author} {\bibfnamefont {J.~A.}\ \bibnamefont
  {Lawlor}}, \bibinfo {author} {\bibfnamefont {P.~D.}\ \bibnamefont {Gorman}},
  \bibinfo {author} {\bibfnamefont {S.~R.}\ \bibnamefont {Power}}, \bibinfo
  {author} {\bibfnamefont {C.~G.}\ \bibnamefont {Bezerra}}, \ and\ \bibinfo
  {author} {\bibfnamefont {M.~S.}\ \bibnamefont {Ferreira}},\ }\href {\doibase
  http://dx.doi.org/10.1016/j.carbon.2014.05.069} {\bibfield  {journal}
  {\bibinfo  {journal} {Carbon}\ }\textbf {\bibinfo {volume} {77}},\ \bibinfo
  {pages} {645 } (\bibinfo {year} {2014})}\BibitemShut {NoStop}%
\bibitem [{\citenamefont {Gorman}\ \emph {et~al.}(2013)\citenamefont {Gorman},
  \citenamefont {Duffy}, \citenamefont {Ferreira},\ and\ \citenamefont
  {Power}}]{Gorman2013}%
  \BibitemOpen
  \bibfield  {author} {\bibinfo {author} {\bibfnamefont {P.~D.}\ \bibnamefont
  {Gorman}}, \bibinfo {author} {\bibfnamefont {J.~M.}\ \bibnamefont {Duffy}},
  \bibinfo {author} {\bibfnamefont {M.~S.}\ \bibnamefont {Ferreira}}, \ and\
  \bibinfo {author} {\bibfnamefont {S.~R.}\ \bibnamefont {Power}},\ }\href
  {\doibase 10.1103/PhysRevB.88.085405} {\bibfield  {journal} {\bibinfo
  {journal} {Phys. Rev. B}\ }\textbf {\bibinfo {volume} {88}},\ \bibinfo
  {pages} {085405} (\bibinfo {year} {2013})}\BibitemShut {NoStop}%
\bibitem [{\citenamefont {Wei{\ss}e}\ \emph {et~al.}(2006)\citenamefont
  {Wei{\ss}e}, \citenamefont {Wellein}, \citenamefont {Alvermann},\ and\
  \citenamefont {Fehske}}]{Weisse2006}%
  \BibitemOpen
  \bibfield  {author} {\bibinfo {author} {\bibfnamefont {A.}~\bibnamefont
  {Wei{\ss}e}}, \bibinfo {author} {\bibfnamefont {G.}~\bibnamefont {Wellein}},
  \bibinfo {author} {\bibfnamefont {A.}~\bibnamefont {Alvermann}}, \ and\
  \bibinfo {author} {\bibfnamefont {H.}~\bibnamefont {Fehske}},\ }\href@noop {}
  {\bibfield  {journal} {\bibinfo  {journal} {Rev. Mod. Phys.}\ }\textbf
  {\bibinfo {volume} {78}},\ \bibinfo {pages} {275} (\bibinfo {year}
  {2006})}\BibitemShut {NoStop}%
\bibitem [{\citenamefont {Covaci}\ \emph {et~al.}(2010)\citenamefont {Covaci},
  \citenamefont {Peeters},\ and\ \citenamefont {Berciu}}]{Covaci2010}%
  \BibitemOpen
  \bibfield  {author} {\bibinfo {author} {\bibfnamefont {L.}~\bibnamefont
  {Covaci}}, \bibinfo {author} {\bibfnamefont {F.}~\bibnamefont {Peeters}}, \
  and\ \bibinfo {author} {\bibfnamefont {M.}~\bibnamefont {Berciu}},\
  }\href@noop {} {\bibfield  {journal} {\bibinfo  {journal} {Phys. Rev. Lett.}\
  }\textbf {\bibinfo {volume} {105}},\ \bibinfo {pages} {167006} (\bibinfo
  {year} {2010})}\BibitemShut {NoStop}%
\bibitem [{\citenamefont {Yuan}\ \emph {et~al.}(2010)\citenamefont {Yuan},
  \citenamefont {De~Raedt},\ and\ \citenamefont {Katsnelson}}]{Yuan2010}%
  \BibitemOpen
  \bibfield  {author} {\bibinfo {author} {\bibfnamefont {S.}~\bibnamefont
  {Yuan}}, \bibinfo {author} {\bibfnamefont {H.}~\bibnamefont {De~Raedt}}, \
  and\ \bibinfo {author} {\bibfnamefont {M.~I.}\ \bibnamefont {Katsnelson}},\
  }\href@noop {} {\bibfield  {journal} {\bibinfo  {journal} {Phys. Rev. B}\
  }\textbf {\bibinfo {volume} {82}},\ \bibinfo {pages} {115448} (\bibinfo
  {year} {2010})}\BibitemShut {NoStop}%
\bibitem [{\citenamefont {Pereira}\ \emph {et~al.}(2008)\citenamefont
  {Pereira}, \citenamefont {Lopes~dos Santos},\ and\ \citenamefont
  {Castro~Neto}}]{Pereira2008}%
  \BibitemOpen
  \bibfield  {author} {\bibinfo {author} {\bibfnamefont {V.~M.}\ \bibnamefont
  {Pereira}}, \bibinfo {author} {\bibfnamefont {J.}~\bibnamefont {Lopes~dos
  Santos}}, \ and\ \bibinfo {author} {\bibfnamefont {A.~H.}\ \bibnamefont
  {Castro~Neto}},\ }\href@noop {} {\bibfield  {journal} {\bibinfo  {journal}
  {Physical Review B}\ }\textbf {\bibinfo {volume} {77}},\ \bibinfo {pages}
  {115109} (\bibinfo {year} {2008})}\BibitemShut {NoStop}%
\bibitem [{\citenamefont {Cheianov}\ \emph {et~al.}(2010)\citenamefont
  {Cheianov}, \citenamefont {Sylju{\aa}sen}, \citenamefont {Altshuler},\ and\
  \citenamefont {Fal'ko}}]{Cheianov2010}%
  \BibitemOpen
  \bibfield  {author} {\bibinfo {author} {\bibfnamefont {V.~V.}\ \bibnamefont
  {Cheianov}}, \bibinfo {author} {\bibfnamefont {O.}~\bibnamefont
  {Sylju{\aa}sen}}, \bibinfo {author} {\bibfnamefont {B.~L.}\ \bibnamefont
  {Altshuler}}, \ and\ \bibinfo {author} {\bibfnamefont {V.~I.}\ \bibnamefont
  {Fal'ko}},\ }\href {http://stacks.iop.org/0295-5075/89/i=5/a=56003}
  {\bibfield  {journal} {\bibinfo  {journal} {EPL (Europhysics Letters)}\
  }\textbf {\bibinfo {volume} {89}},\ \bibinfo {pages} {56003} (\bibinfo {year}
  {2010})}\BibitemShut {NoStop}%
\bibitem [{\citenamefont {Abanin}\ \emph {et~al.}(2010)\citenamefont {Abanin},
  \citenamefont {Shytov},\ and\ \citenamefont {Levitov}}]{Abanin2010}%
  \BibitemOpen
  \bibfield  {author} {\bibinfo {author} {\bibfnamefont {D.~A.}\ \bibnamefont
  {Abanin}}, \bibinfo {author} {\bibfnamefont {A.~V.}\ \bibnamefont {Shytov}},
  \ and\ \bibinfo {author} {\bibfnamefont {L.~S.}\ \bibnamefont {Levitov}},\
  }\href {\doibase 10.1103/PhysRevLett.105.086802} {\bibfield  {journal}
  {\bibinfo  {journal} {Phys. Rev. Lett.}\ }\textbf {\bibinfo {volume} {105}},\
  \bibinfo {pages} {086802} (\bibinfo {year} {2010})}\BibitemShut {NoStop}%
\bibitem [{\citenamefont {{Santos}}\ and\ \citenamefont
  {{Henrard}}(2014)}]{Santos2014}%
  \BibitemOpen
  \bibfield  {author} {\bibinfo {author} {\bibfnamefont {H.}~\bibnamefont
  {{Santos}}}\ and\ \bibinfo {author} {\bibfnamefont {L.}~\bibnamefont
  {{Henrard}}},\ }\href@noop {} {\bibfield  {journal} {\bibinfo  {journal}
  {ArXiv e-prints}\ } (\bibinfo {year} {2014})},\ \Eprint
  {http://arxiv.org/abs/1405.4911} {arXiv:1405.4911 [cond-mat.mtrl-sci]}
  \BibitemShut {NoStop}%
\bibitem [{\citenamefont {Li}\ and\ \citenamefont {Thouless}(1989)}]{Li1989}%
  \BibitemOpen
  \bibfield  {author} {\bibinfo {author} {\bibfnamefont {Q.}~\bibnamefont
  {Li}}\ and\ \bibinfo {author} {\bibfnamefont {D.}~\bibnamefont {Thouless}},\
  }\href@noop {} {\bibfield  {journal} {\bibinfo  {journal} {Phys. Rev. B}\
  }\textbf {\bibinfo {volume} {40}},\ \bibinfo {pages} {9738} (\bibinfo {year}
  {1989})}\BibitemShut {NoStop}%
\bibitem [{\citenamefont {Zhu}(2014)}]{JunPrivComm}%
  \BibitemOpen
  \bibfield  {author} {\bibinfo {author} {\bibfnamefont {J.}~\bibnamefont
  {Zhu}},\ }\href@noop {} {}\bibinfo {howpublished} {private communication}
  (\bibinfo {year} {2014})\BibitemShut {NoStop}%
\end{thebibliography}

%

\end{document}